\documentclass[epj,final]{svjour}
\usepackage{graphicx,amsmath}

\begin{document}

\title{Infrared and electronic Raman response of coexisting $d$-wave density wave and $d$-wave superconductivity}
\titlerunning{Infrared and Raman response of coexisting $d$-density wave and $d$-wave superconductivity}

\author{
Andr\'as V\'anyolos\inst{1}\fnmsep\thanks{E-mail: \texttt{vanyolos@kapica.phy.bme.hu}}
\and Bal\'azs D\'ora\inst{1}
\and Attila Virosztek\inst{1,2}}
\authorrunning{Andr\'as V\'anyolos et al.}

\institute{
Department of Physics, Budapest University of Technology and Economics, 1521 Budapest, Hungary
\and
Research Institute for Solid State Physics and Optics, PO Box 49, 1525 Budapest, Hungary}

\date{Received: date / Revised version: date}

\abstract{We present mean-field calculations for the in-plane optical conductivity, the superfluid density, and the electronic
  Raman susceptibility in quasi two-dimensional systems possessing a ground state with two competing order parameters: $d$-wave
  density wave (dDW) and $d$-wave superconductor (dSC). In the coexisting dDW+dSC phase we calculate the frequency dependence of
  these correlation functions in the presence of impurity scattering in the unitary limit, relevant to zinc-doped cuprate
  superconductors.}

%\pacs{74.20.-z, 74.25.Gz, 78.30.Ly}
\maketitle

\section{Introduction}\label{sec:intro}
The pseudogap phase of underdoped high-temperature superconductors (HTSC) has attracted considerable theoretical
\cite{laughlin,emery,benfatto,zeyher-greco,valenzuela} and experimental \cite{ding,loeser,norman,renner} attention over recent
years. Despite numerous efforts the microscopic origin of the pseudogap and in particular its relationship with superconductivity
still remains an open question. The destruction of the Fermi surface above the superconducting transition temperature $T_c$ has
been detected by a series of experimental techniques, both indirect and direct, including tunneling \cite{renner} and
angle-resolved photoemission spectroscopy \cite{ding,loeser,norman}.

As to the theoretical description of this striking phenomenon, one class of models attributes the pseudogap phase as a precursor
of superconductivity in which Cooper pairs form at a temperature $T^*$ but only acquire phase coherence at a lower temperature
$T_c$, where they form a uniform $d$-wave superconducting (dSC) condensate \cite{emery}. Within this framework the pseudogap and
the superconducting gap are intimately related. As opposed to this, another class of models invokes various phases which are not
directly related to superconductivity, but rather compete with it. One of these is the $d$-wave density wave (dDW) scenario
\cite{laughlin}, which has earned particular interest by now. The dDW is a prototype of an unconventional density wave (UDW), a
particle-hole condensate with a fully developed gap of $d$-wave symmetry. Its most interesting property is perhaps the fact that
it lacks electronic charge density modulation, hence the name ``hidden order'', and that there are real staggered orbital currents
circulating in the plaquettes of the CuO$_2$ plane in an alternating fashion. Much attention has been devoted to show the
consistency of the dDW scenario with several properties of HTSC. These include transport properties such as the in-plane and
$c$-axis optical conductivities \cite{valenzuela,kim}, electronic Raman response \cite{valenzuela}, thermodynamic properties
\cite{congjun1}, and quasiparticle scattering on a localized impurity \cite{bena,morr,jian}, just to mention a few. Although the
dDW scenario was introduced initially on phenomenological grounds \cite{laughlin}, its possible occurrence in microscopic models
like the half-filled two-leg ladder has been checked since then \cite{marston}.

As we have mentioned so far, the pure dDW phase has been studied quite extensively in recent years from many aspects due to its
potential applicability to the anomalous normal state of underdoped cuprates. However, the possible coexistence of this
particle-hole condensate with $d$-wave superconductivity below $T_c$ and the consequences of the two-gap feature has not been
fully developed, at least as far as the frequency dependent correlation functions are concerned
\cite{zeyher-greco,ismer-spin}. Mean-field calculations show that there is a substantial region on the temperature-doping phase
diagram of the cuprates where the energetically stable solution of the gap equations is a mixture of dDW and dSC orders
\cite{zeyher-greco,jian,ghosh-coexistance}. This coexisting dDW+dSC state has been studied within the context of the $c$-axis
optical sum rule and superfluid density of HTSC \cite{kim}. Also, there has been considerable interest devoted to the exploration
of quasiparticle interference patterns and local density of states around a single impurity \cite{bena,jian,andrenacci}.

Partly motivated by recent infrared and electronic Raman scattering measurements showing evidence for two-gap feature in
underdoped HTSC \cite{gallais,yu}, the aim of this paper is to go beyond the study of single particle properties of dDW+dSC and
calculate two-particle Green's functions. Namely, we present analytical calculations of the in-plane optical conductivity and the
electronic Raman susceptibility of dDW+dSC. Disorder effects are included within the unitary limit of impurity scattering relevant
to Zn-doped high-$T_c$ cuprates \cite{pan,hotta}. Also, coherence effects are studied in detail: we find that while
electromagnetic absorption (conductivity) is a clear type I process in pure (single component) dDW, electronic Raman response can
be either type I or II, depending on the symmetry of the Raman vertex. In particular we find that in $A_{1g}$ and $B_{1g}$
channels the dDW coherence factor is type I, whereas interestingly in $B_{2g}$ it changes to type II. This Raman example shows
that one has to be rather careful in determining the coherence factor of a response function in the different condensates,
i.e.\ in dDW and dSC, because the general rule of thumb according to which one only has to change the coherence factor from type I
to II (or vice versa) when switching between dDW and dSC, might not work every time. It works for example in the case of infrared
response, but not necessarily in the Raman experiment. Further examples are given in Section~\ref{sec:infrared}.

The paper is organized as follows. In Sec.~\ref{sec:formalism} we describe our model and the analytic formalism. In
Sec.~\ref{sec:infrared} we calculate the optical conductivity and the superfluid density in dDW+dSC, while in Sec.~\ref{sec:raman}
electronic Raman scattering is studied in the relevant $A_{1g}$, $B_{1g}$, and $B_{2g}$ symmetries. Finally, in
Sec.~\ref{sec:conclusions} we conclude and summarize. An Appendix is also given for some technical details.

\section{Formalism}\label{sec:formalism}

Let us consider the following effective low-energy Hamiltonian \cite{pssb-gossamer,pssb-bottomup,pssb-hightc}
\begin{align}
H_0&=\sum_{\mathbf{k},\sigma}(\epsilon(\mathbf{k})-\mu)c^+_{\mathbf{k},\sigma}c_{\mathbf{k},\sigma}\notag\\
&\phantom{=}+\sum_{\mathbf{k},\sigma}\left(i\Delta_1(\mathbf{k})c^+_{\mathbf{k},\sigma}c_{\mathbf{k+Q},\sigma}+\text{h.c.}\right)\notag\\
&\phantom{=}+\sum_{\mathbf{k}}\left(\Delta_2(\mathbf{k})c^+_{\mathbf{k},\uparrow}c^+_{-\mathbf{k},\downarrow}+\text{h.c.}\right),\label{h0}
\end{align}
where $\epsilon(\mathbf{k})$ is the electronic kinetic energy, $\mu$ is the chemical potential, $\mathbf{Q}=(\pi/a,\pi/a)$ is the
dDW nesting vector, and the amplitudes and momentum dependent parts of the dDW and dSC order parameters, respectively, separate
via $\Delta_1(\mathbf{k})=\Delta_1f(\mathbf{k})$ and $\Delta_2(\mathbf{k})=\Delta_2f(\mathbf{k})$, with
\begin{equation}
f(\mathbf{k})=(\cos(ak_x)-\cos(ak_y))/2.\label{gapf}
\end{equation}
Now the Green's function of the pure system without impurities is obtained as \cite{pssb-bottomup,pssb-hightc}
\begin{multline}
G_0(\mathbf{k},i\omega_n)^{-1}=i\omega_n-\xi(\mathbf{k})\rho_3\sigma_3+\eta(\mathbf{k})\sigma_3\\
+\Delta_1f(\mathbf{k})\rho_2-|\Delta_2|f(\mathbf{k})e^{i\phi_2\sigma_3}\rho_3\sigma_1,
\label{g0}
\end{multline}
where $\xi(\mathbf{k})=(\epsilon(\mathbf{k})-\epsilon(\mathbf{k+Q}))/2$, the imperfect nesting term
\cite{pssb-gossamer,pssb-bottomup,pssb-hightc} reads
\begin{equation}
\eta(\mathbf{k})=\mu-(\epsilon(\mathbf{k})+\epsilon(\mathbf{k+Q}))/2,\label{impnest}
\end{equation}
and the corresponding four-dimensional spinor field is defined as
\begin{equation}
\Psi^+(\mathbf{k})=(c^+_{\mathbf{k},\uparrow},c_{-\mathbf{k},\downarrow},c^+_{\mathbf{k+Q},\uparrow},
c_{-\mathbf{k}-\mathbf{Q},\downarrow}).\label{spinor}
\end{equation}
As is well known from dDW theory \cite{laughlin,valenzuela,andrenacci}, the order parameter of dDW is purely imaginary. Thus, by
factoring out the imaginary unit $\Delta_1$ must be necessarily real. To keep the following calculations as simple as possible,
but without the loss of generality we may choose the dSC order parameter $\Delta_2$ to be real too, because its phase $\phi_2$ is
unrestricted in our mean-field theory. From the poles of $G_0(\mathbf{k},\omega)$ the quasiparticle excitation spectrum is
obtained as $\omega=\pm E_\pm(\mathbf{k})$, with
\begin{equation}
E_\pm(\mathbf{k})=\sqrt{\left(\pm\sqrt{\xi^2(\mathbf{k})+\Delta_1^2(\mathbf{k})}-\eta(\mathbf{k})\right)^2
+\Delta_2^2(\mathbf{k})}.\label{generalpole}
\end{equation}

\begin{figure}[t]
\begin{center}
\includegraphics[width=0.9\columnwidth]{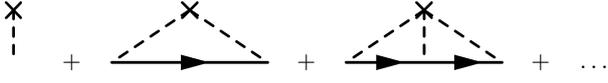}
\end{center}
\caption{Infinite series of non-crossing self-energy (or $T$-matrix) diagrams due to impurity scattering. The solid line
  denotes the electron propagator $G$, while the dashed line is for electron-impurity interaction. Dashed lines coming from
  the same cross represent successive scattering on the same impurity.\label{fig:selfenergy}}
\end{figure}

Impurity scattering can be incorporated into the theory along the lines developed first by Abrikosov and Gor\-kov
in the early sixties \cite{abrikosov,skalski,ambegaokar}. The interaction of electrons with non-magnetic point-like
impurities is described by the Hamiltonian \cite{balazs-born,balazs-unitary,nca-impurity}
\begin{equation}
H_1=\frac1V\sum_{\mathbf{k,q},j}e^{-i\mathbf{qR}_j}\Psi^+(\mathbf{k+q})U(\mathbf{R}_j)
\Psi(\mathbf{k}),\label{h1}
\end{equation}
where the potential in the four-dimensional Nambu-Gorkov space reads
\begin{equation}
U(\mathbf{R}_j)=U
\begin{pmatrix}
1 & 0 & e^{i\mathbf{QR}_j} & 0\\
0 & -1 & 0 & -e^{i\mathbf{QR}_j}\\
e^{-i\mathbf{QR}_j} & 0 & 1 & 0\\
0 & -e^{-i\mathbf{QR}_j} & 0 & -1
\end{pmatrix},\label{potential}
\end{equation}
with $\mathbf{R}_j$ being the position of the $j$th impurity atom and $U$ is the amplitude of the delta-potential. Following the
method of Ref.~\cite{balazs-born}, the electronic self-energy from impurities is
\begin{subequations}\label{coupled}
\begin{equation}
\Sigma_\mathbf{R}(i\omega_n)=n_\text{i}\left(U(\mathbf{R})^{-1}-\frac1V\sum_\mathbf{k}G(\mathbf{k},i\omega_n)\right)^{-1},
\label{selfenergy}
\end{equation}
where $n_\text{i}=N_\text{i}/V$ is the impurity concentration and the subscript $\mathbf{R}$ means the position of an impurity
over which the average should be taken. Its diagrammatic representation is shown in Fig.~\ref{fig:selfenergy}. But most
importantly, because of the requirement of self-consistency, $G(\mathbf{k},i\omega_n)$ is itself the renormalized Green's function
appearing in Dyson's equation
\begin{equation}
G(\mathbf{k},i\omega_n)^{-1}=G_0(\mathbf{k},i\omega_n)^{-1}-\langle\Sigma_\mathbf{R}(i\omega_n)\rangle_\text{imp}.
\label{dyson}
\end{equation}
\end{subequations}

Before proceeding further with the analytical calculation we have to comment on an important simplification we will employ. The
momentum dependence of the quasiparticle band structure given by Eq.~\eqref{generalpole} is very complicated because of the finite
imperfect nesting term. As is well known by now from ARPES measurements underdoped cuprates do not exhibit perfect nesting and the
experimental situation is best described by a $t$-$t'$ model, where both the next-nearest neighbor hopping term $t'$ and the
doping $\mu$ is finite \cite{ismer-spin,andrenacci}. This gives rise to finite $\eta$, and consequently to a rich
temperature-doping phase diagram \cite{jian,ghosh-coexistance}. However, in our model calculation focusing on the coherence
effects coming from the different components of the dDW+dSC condensate, in a first approximation it is sufficient to study the
$\eta=0$ case only. As we shall see shortly, the effect of coherence factors on the lineshapes of the dynamical susceptibilities
are rather robust and remain strong in a system with imperfect nesting. An inclusion of a finite but small $\eta(\mathbf{k})$ would of
course introduce additional features in the spectra, but this would require a fully numerical approach and is therefore out of the
scope of the present investigation. Nevertheless, further developments in this direction are highly desired.

\unitlength 0.35mm

Neglecting imperfect nesting we find that electron or hole pockets will not form around the dDW gap nodes and the two gap
functions add up squared as the quasiparticle energies become doubly degenerate
\begin{equation}
E_\pm(\mathbf{k})=E(\mathbf{k})=\sqrt{\xi^2(\mathbf{k})+(\Delta_1^2+\Delta_2^2)f^2(\mathbf{k})}.
\label{specpole}
\end{equation}
It is clear that these quasiparticles behave in some respect the same as those in dSC. Now, however, they feel a combined
effective $d$-wave gap $\Delta(\mathbf{k})=\Delta f(\mathbf{k})$, where
\begin{equation}
\Delta=\sqrt{\Delta_1^2+\Delta_2^2}.\label{effectivegap}
\end{equation}
This mapping between a dSC and the particle-hole symmetric dDW+dSC involves that the whole thermodynamic behavior of the latter,
as well as its response to static external perturbations are the same as those of a dSC \cite{maki-dwave,maki-nodal}, because the
density of states (DOS) is itself the same. Only the energy scale is different according to Eq.~\eqref{effectivegap}. By direct
substitution one can also verify that in unitary limit, where $U\to\infty$, the solution of Eqs.~\eqref{coupled} is
\begin{multline}
G(\mathbf{k},i\omega_n)^{-1}=i\tilde\omega_n-\xi(\mathbf{k})\rho_3\sigma_3\\
+\Delta_1f(\mathbf{k})\rho_2-\Delta_2f(\mathbf{k})\rho_3\sigma_1,\label{g}
\end{multline}
where the Matsubara frequency is the only quantity to be renormalized by impurities. The detailed analysis of the unitary limit
is covered in many textbooks \cite{maki-dwave,maki-nodal,abrikosov-book,rickayzen-book,parks-book} and
publications \cite{balazs-born,balazs-unitary,sun-maki,puch-maki}, therefore we shall not repeat it here. We shall only summarize
some essential expressions that will be utilized repeatedly during the calculation of the complex conductivity and the Raman
response. The analytically continued renormalized frequency $u=i\tilde\omega(i\omega_n\to\omega+i0)/\Delta$ obeys
\begin{equation}
\frac{\omega}{\Delta}=u+\alpha\frac{\pi}{2}\frac{\sqrt{1-u^2}}{uK\left(\frac{1}{\sqrt{1-u^2}}\right)},\label{unitary}
\end{equation}
where $K(z)$ is the complete elliptic integral of the first kind, $\alpha=\Gamma/\Delta$ is the dimensionless pair-breaking
parameter, $\Gamma=2n_\text{i}/(\pi N_0)$ is the scattering rate, and $N_0$ is the normal state density of states per spin. In
arriving to Eq.~\eqref{unitary} the momentum integrals were performed analytically in the continuum limit, where the Fermi surface
is assumed to be cylindrical, the $d$-wave angular dependence simplifies to $f(\mathbf{k})=\cos(2\phi)$, and thus the Fermi
surface average is simply an average over $\phi$ \cite{maki-dwave,maki-nodal,jiang-inelastic}. We note that this commonly used
simplification of momentum integrals is purely technical. It allows for analytical calculations, but at the same time captures the
essence of physics in the sense that the qualitative features remain the same as those obtained from numerical integration with
the exact lattice periodic functions \cite{valenzuela,devereaux-clean}. The renormalized Matsubara frequency at zero external
frequency, namely $C_0=\tilde\omega(\omega_n=0)$, plays a central role in the theory \cite{inplane-dsc,outofplane-dsc}. It is
given by
\begin{equation}
C_0^2=\alpha\frac{\pi}{2}\sqrt{1+C_0^2}\left[K\left(\frac{1}{\sqrt{1+C_0^2}}\right)\right]^{-1}.\label{c0unitary}
\end{equation}

Apropos of Eq.~\eqref{effectivegap} we have already pointed out that the single particle properties of a dDW+dSC state with $\eta$
set to zero are the same as those of a dSC due to the matching DOS \cite{maki-dwave,inplane-dsc}. The similarities, however, end
here because at finite frequency the dynamical response is dominated by coherence effects. These reveal the true nature of the
ground state, the fact that it is actually a competing superposition of dDW and dSC condensates, both with its own order
parameter.

\section{Optical conductivity}\label{sec:infrared}

In this section we calculate the frequency dependence of the in-plane optical conductivity and the superconducting condensate
fraction at zero temperature for the coexisting dDW+dSC phase with impurities. Following standard methods
\cite{skalski,rickayzen-book,mahan-book} the complex conductivity is obtained from the correlation function
$\chi=\langle[J_\alpha,J_\beta]\rangle$, where the paramagnetic current operator reads in our four-dimen\-sional Nambu notation as
\begin{equation}
J_\alpha(\mathbf{q})=-e\sum_\mathbf{k}v_\alpha(\mathbf{k}+\mathbf{q}/2)\Psi^+(\mathbf{k})\rho_3\Psi(\mathbf{k+q}).
\label{current}
\end{equation}
In this expression it is the $\rho_3$ matrix structure that should be really kept in mind, because this is the essential
ingredient here that distinguishes the infrared response from the Raman response via the different coherence factors. As to the
specific momentum dependence of the quasiparticle velocity, in accord with our assumption of a cylindrical Fermi surface, we use
$v_\alpha(\mathbf{k})=k_\alpha/m$, with $m$ being the effective mass of the particles. With this the current correlator is
\begin{multline}
\chi(i\nu_n)=-\delta_{\alpha\beta}\frac{4\pi e^2n}{m^2}\frac{T}{V}\\
\times\sum_{\mathbf{k},\omega_n}
\frac{-\tilde\omega_n\tilde\omega_n'+\xi^2(\mathbf{k})-\Delta_1^2(\mathbf{k})+\Delta_2^2(\mathbf{k})}
{\left(\tilde\omega_n^2+E^2(\mathbf{k})\right)\left(\tilde\omega_n'^2+E^2(\mathbf{k})\right)},
\label{condcoherence}
\end{multline}
where $n$ is the total electron density, and the prime on $\tilde\omega_n'$ means that the renormalized Matsubara frequency has to
be computed at the shifted frequency $\omega_n+\nu_n$.

At this point it is worth stopping for a moment and discuss coherence effects in the different condensates. In BCS theory
\cite{parks-book,tinkham-book} the coherence factors show up in the calculation of two-particle Green's functions, a typical
example of which is the current-current correlator we are just studying in Eq.~\eqref{condcoherence}. In pure superconductors the
frequency sum can be performed \cite{mahan-book,fetter-book}, leading to the appearance of the coherence factors. For brevity we
only quote here the long wavelength result associated with the creation of two quasiparticles, it is:
$1-(\xi^2\mp\Delta^2)/E^2$. The upper and lower signs refer to type I and type II coherence factors, respectively. Electromagnetic
absorption is a representative of type II processes, where the factor vanishes exactly, signaling the well known fact that at
finite frequency the optical response of a pure superconductor is zero.\footnote{Note, however, that though the real part of the
  conductivity vanishes at every finite frequency, the conductivity sum rule requires the optical weight to be conserved, which
  means that all spectral weight is necessarily shifted to $\omega=0$ in the form of a Drude peak $(D\delta(\omega))$.} Coming
back to our original formula in Eq.~\eqref{condcoherence} we see that due to the presence of impurities the Matsubara sum cannot
be done analytically anymore, instead, it is the momentum integration that will be performed first. Nevertheless, it is not too
difficult to figure out even from this expression that the plus sign in front of the superconducting order parameter would involve
clear type II coherence effects if the ground state was a single component dSC. As opposed to this in a single component dDW,
where only $\Delta_1(\mathbf{k})$ exists, the sign is just the opposite, leading to clear type I behavior. From this example one
might have the impression that the same physical quantity is always of different type in a superconductor and a density
wave. Another example of great importance is the charge density with $\sigma_3$ matrix structure, which has plus and minus signs
in front of the dDW and dSC order parameters, respectively, i.e.\ the situation is just the opposite compared to the charge
current. This general rule of thumb seems to work usually, and is related to the different structures of the Nambu space in these
condensates. In density waves it is the $\mathbf{k}\to\mathbf{k+Q}$ transformation that spans the space, and the spinor $\Psi$
consists of particle type operators only.  In superconductors, on the other hand, the transformation properties of the physical
quantities are studied under inversion in momentum space, that is under $\mathbf{k}\to-\mathbf{k}$, and the spinor mixes particle
and hole type operators. This extra canonical transformation from particles to holes introduces an additional minus sign compared
to that in a density wave. But there are interesting exceptions as well. One of these is the electronic Raman susceptibility,
which, as we shall see in Section~\ref{sec:raman}, turns out to be a type I process in superconductors. In dDW, however, the
symmetry of the Raman vertex decides whether it is type I or type II. Another exception is the spin density whose matrix structure
is simply the identity, leading to plus signs in the numerator and type II behavior in both condensates. In order to give a brief
and easy-to-use reference of all these results we gathered these physical quantities and the appropriate coherence factors in
Table~\ref{table:coh}, with special emphasis on the difference between the dDW and dSC cases.

\begin{table}[t]
\begin{tabular*}{\columnwidth}{l@{\extracolsep\fill}c@{\extracolsep\fill}c@{\extracolsep\fill}c}
\hline\hline
Physical quantity & Nambu structure & dDW & dSC\\
\hline
Charge current & $\rho_3$ & $-$ & $+$\\
Charge density & $\sigma_3$ & $+$ & $-$\\
Spin density & $1$ & $+$ & $+$\\
Raman $A_{1g}$ & $\rho_3\sigma_3$ & $-$ & $-$\\
Raman $B_{1g}$ & $\rho_3\sigma_3$ & $-$ & $-$\\
Raman $B_{2g}$ & $\sigma_3$ & $+$ & $-$\\
\hline\hline
\end{tabular*}
\caption{\label{table:coh}Coherence factors associated with a few physical quantities in dDW and dSC condensates. The $\pm$ signs
  appear in front of the order parameters in the one bubble calculation of the relevant correlation functions. For examples see
  Eqs.~\eqref{condcoherence} and \eqref{ramancoherence}. The minus (plus) sign is commonly referred to as a type I (II) coherence
  factor \cite{tinkham-book}.}
\end{table}

Now, after carrying out the momentum integration in Eq.~\eqref{condcoherence} and transforming the frequency sum into a contour
integral the (regular) real part of the optical conductivity can be expressed as \cite{balazs-born,balazs-unitary,inplane-dsc}
\begin{equation}
\sigma_1(\omega)=-\frac{e^2n}{m\pi\Delta}\frac{I_\text{n}(\omega)+I_\text{pb}(\omega)}{\omega},
\label{sigma1}
\end{equation}
where, using density wave terminology \cite{impurity-sdw}
\begin{subequations}\label{integrals}
\begin{multline}
I_\text{n}(\omega)=\int_0^\infty\text{d}x\left(\tanh\left(\frac{x+\omega}{2T}\right)-
\tanh\left(\frac{x}{2T}\right)\right)\\
\times\text{Re}\left\{F(u(x+\omega),u(x))-F(u(x+\omega),\overline{u(x)})\right\}\label{in}
\end{multline}
is the so-called normal (intraband), while
\begin{multline}
I_\text{pb}(\omega)=\int_0^\omega\text{d}x\tanh\left(\frac{x}{2T}\right)\\
\times\text{Re}\left\{F(u(x-\omega),u(x))-F(u(x-\omega),\overline{u(x)})\right\}\label{ipb}
\end{multline}
\end{subequations}
is the pair-breaking (interband) contribution. Further, the dimensionless function $F$ is
\begin{multline}
F(u,u')=\frac{1}{{u'}^2-u^2}\Bigg\{-u(u+u')\frac{K}{\sqrt{1-u^2}}\\
+2\frac{\Delta_1^2}{\Delta^2}\left(\sqrt{1-u^2}E+\frac{u^2K}{\sqrt{1-u^2}}\right)-[u\leftrightarrow u']\Bigg\},\label{f}
\end{multline}
where the argument of $K$ and $E$, the complete elliptic integrals of the first and second kind, is $1/\sqrt{1-u^2}$. This is the
generalization of the corresponding $F$ functions of dSC \cite{inplane-dsc} and UDW \cite{balazs-born}. Strictly speaking, as it
stands above, it yields only the quasiparticle contribution to the conductivity. However, due to the fact that in our
approximation the scattering potential $U$ (see Eq.~\eqref{potential}) is momentum independent, the vertex corrections exactly
vanish and thus Eq.~\eqref{f} gives the total contribution to $\sigma_1$. This is not necessarily so in the Raman calculation, as
we shall see shortly in Section~\ref{sec:raman}, because there the ladder diagrams can renormalize the quasiparticle result, for
example in the $B_{1g}$ symmetry, and for this to see the vertex equation has to be solved too.

\begin{figure}[t]
\includegraphics[width=\columnwidth]{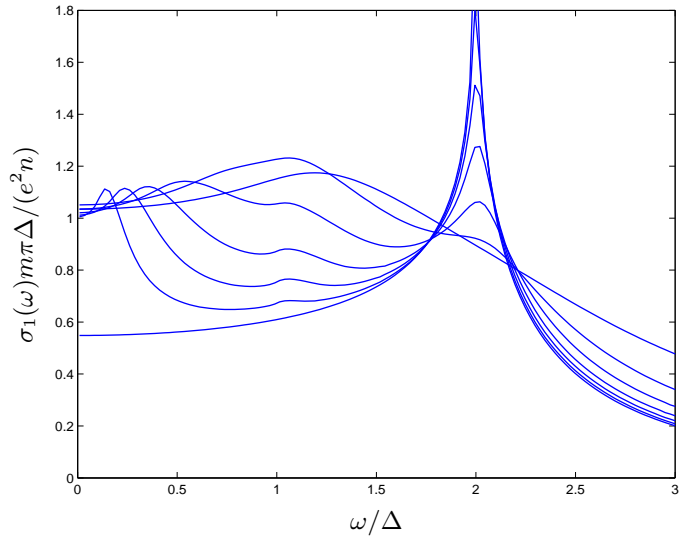}
\caption{\label{fig:sigma1u}(Color online) Zero temperature optical conductivity of dDW+dSC in the unitary limit for $\alpha=0$,
  0.01, 0.025, 0.05, 0.1, 0.2, and 0.5. The curve belonging to $\alpha=0$ meets the vertical axis at $\omega=3\Delta$ at the
  lowest. Further $\Delta_1/\Delta=2/3$.}
\end{figure}

The conductivity at zero temperature in the coexisting dDW+dSC phase is shown in Fig.~\ref{fig:sigma1u} for different values of
impurity concentrations. Note that on different curves the only parameter varying is $\alpha$. The gap amplitudes are fixed, with
the constraint in Eq.~\eqref{effectivegap} of course, and their renormalization due to impurity scattering is not taken here into
account.\footnote{For this we would have to solve the coupled gap equations for $\Delta_1$ and $\Delta_2$ self-consistently, for
  each $\alpha$, but this enterprise is out of the scope of this publication.} Rather, we can think of these as experimentally
adjustable parameters of the model. We can see that the stronger the pair-breaking is, the more the lineshapes approach the
metallic Lorentzian form. One could think, naively, that the robust peak around $\omega=2\Delta$ must be the result of the density
of states, in particular its Van Hove singularities located at the gap edge $\omega=\pm\Delta$ \cite{maki-nodal,inplane-dsc}. As
the direct calculation verifies, this argumentation holds only in unconventional density waves \cite{balazs-born,balazs-unitary},
like the dDW itself \cite{valenzuela}, but in dSC the type II coherence factor suppresses this transition
\cite{inplane-dsc}. Thus, the fact that this strong peak still develops, can be attributed to the dDW component of the ground
state alone. The small bump at $\omega=\Delta$ results from transitions from and to the narrow impurity band around the Fermi
energy \cite{hotta,balazs-unitary}. It is also worth mentioning that in the dc conductivity the $\alpha\to0$ and $\omega\to0$
limits cannot be interchanged, this is why we obtain different result in the pure case. We believe that the right procedure is the
one where the $\omega\to0$ limit is taken first \cite{balazs-unitary}.

The dc conductivity in the ground state is obtained from Eq.~\eqref{sigma1} as
\begin{equation}
\sigma_1(0)=\frac{e^2n}{m\pi\Delta}\frac{1}{\sqrt{1+C_0^2}}E\left(\frac{1}{\sqrt{1+C_0^2}}\right),\label{dc}
\end{equation}
where $C_0$ has been defined previously in Eq.~\eqref{c0unitary}. This is formally the same as the one in dSC \cite{inplane-dsc},
but again $\Delta$ is the amplitude of the effective gap, which is always larger than that of the dSC component, $\Delta_2$. As
the terms containing $C_0$ on the right hand side are always of order unity, and practically do not depend on impurities, this
result for the dc conductivity is a manifestation of Lee's universality relation \cite{lee}.

It might be of interest to examine the pure case analytically too. For $\alpha=0$ we find
\begin{multline}\label{cleancond}
\sigma_1(\omega)=\frac{e^2n}{m\pi\Delta}\left(\frac{\Delta_1}{\Delta}\right)^2\tanh\left(\frac{\omega}{4T}\right)\\
\times\frac{\pi}{2x}
\begin{cases}
x^{-1}\left(K(x)-E(x)\right) & x<1,\\
K(x^{-1})-E(x^{-1}) & x>1,
\end{cases}
\end{multline}
where $x=\omega/(2\Delta)$. Perhaps this is the result that shows most concisely what we have discussed above apropos of
Fig.~\ref{fig:sigma1u}. Namely, that it is the dDW component of the dDW+dSC condensate that leads ultimately to a finite optical
response, and in particular to the logarithmic singularity at the maximal optical gap $2\Delta$. If the ground state consists
entirely of a single component dDW, with $\Delta=\Delta_1$, this expression correctly simplifies to the UDW result
\cite{balazs-sdw}.

The fact that even in the pure case the dDW+dSC state exhibits finite absorption implies the weight of the Drude peak should
decrease. In other words due to the presence of the dDW component spectral weight must be transferred from zero to finite
frequency. This decrease of the Drude weight can be summarized nicely in the form of the in-plane superfluid density
$\rho_\text{s}(T,\Gamma)$, which is calculated at $T=0$ as
\begin{multline}
\rho_\text{s}(0,\Gamma)=\left(1-\frac{\Delta_1^2}{\Delta^2}\right)\\
\times\Bigg(1-\frac{\alpha}{C_0}
+\alpha\int_{C_0}^\infty\frac{\text{d}u}{u^2}\left(1-\frac{E}{K}\right)^2\Bigg).\label{rhounitary}
\end{multline}
The argument of the complete elliptic integrals $K$ and $E$ is $1/\sqrt{1+u^2}$. It is very interesting to observe that what
appears in the large parentheses is formally the condensate fraction of dSC \cite{outofplane-dsc}. More remarkable is, however,
that because of the prefactor not even in the pure system would the superfluid density reach unity
\cite{pssb-gossamer,pssb-bottomup}. The missing electrons are in fact not missing but are frozen in the dDW component of the
ground state. Recall that in a density wave, unless the phase fluctuations of the order parameter are taken into account, there is
no dc conductivity in the ground state \cite{lra,phonon-ucdw}.

\section{Electronic Raman scattering}\label{sec:raman}

In this section we proceed with the calculation of the electronic Raman susceptibility in dDW+dSC with impurities. The calculation
is done along the same lines as the complex conductivity. In this case, however, instead of the current, the correlation function
of the effective density has to be computed, where the velocity vertex is replaced by the Raman vertex
$\gamma(\mathbf{k})$ \cite{dierker,devereaux-rmp}. The operator of the effective density is
\begin{equation}
\tilde\rho(\mathbf{q})=\sum_{\mathbf{k},\sigma}\gamma(\mathbf{k}+\mathbf{q}/2)c_{\mathbf{k},\sigma}^+c_{\mathbf{k+q},\sigma}.
\label{ramanvertexfem}
\end{equation}
In the long wavelength limit the vertex is related to the generalized mass tensor of
electrons \cite{devereaux-clean,tutto-cdw-sc,raman-udw}. We, however, instead of making an ansatz for its momentum dependence,
which obviously depends strongly on the details of the dispersion $\epsilon(\mathbf{k})$, expand it as usual with the basis
functions of the tetragonal point group $D_{4h}$ relevant to cuprate superconductors \cite{jiang-inelastic,sigrist,scalapino}
\begin{equation}
\gamma(\mathbf{k})=\sum_{L,\mu}\gamma_L^\mu\lambda^{\mu}_L(\mathbf{k}).
\label{kifejtes}
\end{equation}
Here $\mu$ represents the even parity irreducible representations of the group, $L$ stands for the $L$th-order contribution to the
vertex which transforms according to the $\mu$th irreducible representation, and the unknown coefficients, responsible for the
overall scale of absorption in the given channel, will be thought of as free parameters to be fitted to experimental data. As to
the basis functions the following low order contributions will be considered \cite{valenzuela,devereaux-clean}
\begin{equation}\label{basisfunctions}
\lambda^{\mu}_L(\mathbf{k})=
\begin{cases}
1,\cos(ak_x)+\cos(ak_y) & \text{$\mu=A_{1g}$, $L=0,4$},\\
\cos(ak_x)-\cos(ak_y) & \text{$\mu=B_{1g}$, $L=2$},\\
\sin(ak_x)\sin(ak_y) & \text{$\mu=B_{2g}$, $L=2$}.
\end{cases}
\end{equation}
It is very interesting to see that among these examples neither the $L=0$ term in the $A_{1g}$ symmetry, which is apparently
nothing else than the density vertex, nor the $B_{2g}$ basis function changes sign under the transformation
$\mathbf{k}\to\mathbf{k+Q}$. This in turn leads to the observation that in Nambu space these contributions will exhibit a matrix
structure equivalent to the charge density, that is $\sigma_3$. In stark contrast to this, all other basis functions do change
sign leading to a distinct $\rho_3\sigma_3$ behavior, and as we shall see in Eq.~\eqref{ramancoherence}, to different coherence
factor as far as the dDW is concerned. With these, our task is now reduced to finding the correlation functions
$\chi^\mu_L=\langle[\tilde\rho^\mu_L,\tilde\rho^\mu_L]\rangle$ of the channel dependent operators
\begin{equation}
\tilde\rho^{\mu}_L(\mathbf{q})=\gamma^{\mu}_L\sum_\mathbf{k}\lambda^{\mu}_L(\mathbf{k})
\Psi^+(\mathbf{k})M\Psi(\mathbf{k+q}),\label{raman}
\end{equation}
where the Nambu matrix $M$ is either $\sigma_3$ or $\rho_3\sigma_3$, depending on the case considered.

\unitlength 0.4mm
\begin{figure}[t]
\begin{center}
\includegraphics[width=0.65\columnwidth]{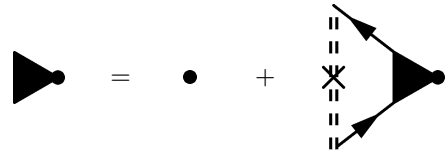}
\end{center}
\caption{Diagrammatic representation of the vertex equation in Eq.~\eqref{vertexeq}. The dot is the bare vertex, the shaded
  triangle represents the vertex correction due to impurities, the cross plays the role of an impurity, the solid line is the
  Green's function, and the double dashed line denotes the self-consistent $T$-matrix, whose diagrams are shown in
  Fig.~\ref{fig:selfenergy}.\label{fig:vertex}}
\end{figure}

Let us start the derivation with the definition of the correlation function we shall compute \cite{devereaux-rmp}. At long
wavelength and imaginary frequency we find
\begin{multline}
\chi^{\mu}_L(i\nu_n)=-(\gamma^{\mu}_L)^2\frac{T}{V}\sum_{\mathbf{k},\omega_n}\lambda^{\mu}_L(\mathbf{k})
\text{Tr}(MG(\mathbf{k},i\omega_n)\\
\times\Lambda^{\mu}_L(\mathbf{k},i\omega_n,i\omega_n+i\nu_n)G(\mathbf{k},i\omega_n+i\nu_n)),\label{def}
\end{multline}
where $G$ is the Green's function from Eq.~\eqref{g}, and for our approximation to be conserving \cite{mahan-book}, the impurity
renormalized vertex $\Lambda^\mu_L$ has to obey the following vertex equation
\begin{multline}
\Lambda^{\mu}_L(\mathbf{k},i\omega_n,i\omega_n+i\nu_n)=\lambda^{\mu}_L(\mathbf{k})M\\
+\frac{n_\text{i}}{V}\sum_\mathbf{p}\frac{1}{N}\sum_\mathbf{R}T_\mathbf{R}(i\omega_n)G(\mathbf{p},i\omega_n)
\Lambda^{\mu}_L(\mathbf{p},i\omega_n,i\omega_n+i\nu_n)\\ \times G(\mathbf{p},i\omega_n+i\nu_n)T_\mathbf{R}(i\omega_n+i\nu_n).
\label{vertexeq}
\end{multline}
Its diagrammatic representation is shown in Fig.~\ref{fig:vertex}. In the full non-crossing approximation the $T$-matrix
incorporates the effect of impurity potential to infinite order. Its diagrams are the same as those of the self-energy in
Fig.~\ref{fig:selfenergy}. Correspondingly, its analytical expression is given by Eq.~\eqref{selfenergy} too, only the factor
$n_\text{i}$ needs to be ignored. In the unitary limit, where $U\to\infty$, it simplifies to
\begin{equation}
T_\mathbf{R}(i\omega_n)=-\left(\sum_\mathbf{k}G(\mathbf{k},i\omega_n)\right)^{-1}.\label{tunitary}
\end{equation}
Now, making use of these results and the fluctuation-dissipation theorem, the Raman intensity is found to be proportional to the
imaginary part of the retarded correlator
\begin{equation}
\text{Im}\chi^{\mu}_L(\omega)=-\frac{2N_0(\gamma^{\mu}_L)^2}{\pi\Delta}
\left(I_\text{n}(\omega)+I_\text{pb}(\omega)\right),\label{ramansusc}
\end{equation}
where the normal and pair-breaking integrals are given by Eqs.~\eqref{integrals}, but now the $F$ function becomes channel
dependent and needs to be calculated for every symmetry separately.\footnote{The conductivity and Raman calculations suggest that
  within mean-field theory every correlation function can be expressed with these two integrals in the long wavelength limit. This
  statement is indeed true, and it follows at once that it is the $F$ function alone that carries all information about the
  specific physical quantity, its matrix structure in Nambu space, the coherence factors involved, and the momentum dependence of
  its kernel. Thus it is obvious that $F$ has to be recomputed for each physical quantity.} These cases will be studied in the
following subsections.

Before turning our attention to the channel dependent spectra and the full solution of the vertex equation, it is worth writing
down the quasiparticle contribution to the susceptibility. Diagrammatically this is the one bubble result, the only term that
survives in a pure system, and its analytical expression is very similar to that of the current-current correlator in
Eq.~\eqref{condcoherence}. It is obtained by approximating the full vertex by the bare one
\begin{multline}
\chi^{\mu}_L(i\nu_n)_\text{qp}=-4(\gamma^{\mu}_L)^2\frac{T}{V}\\
\times\sum_{\mathbf{k},\omega_n}\lambda^{\mu}_L(\mathbf{k})^2
\frac{-\tilde\omega_n\tilde\omega_n'+\xi^2(\mathbf{k})\pm\Delta_1^2(\mathbf{k})-\Delta_2^2(\mathbf{k})}
{\left(\tilde\omega_n^2+E^2(\mathbf{k})\right)\left(\tilde\omega_n'^2+E^2(\mathbf{k})\right)},
\label{ramancoherence}
\end{multline}
where the plus and minus signs in front of the dDW gap belong to $M=\sigma_3$ and $M=\rho_3\sigma_3$, respectively. From this we
can see explicitly what we have already indicated before, that (i) the appearance of $\sigma_3$ leads indeed to a sign change in
front of the superconducting gap compared to Eq.~\eqref{condcoherence}, turning the Raman scattering to a type I process in
superconductors, and (ii) in dDW it depends on the symmetry of the vertex whether inelastic light scattering is a type I or II
process, see Table~\ref{table:coh}.

\subsection{Raman spectra in $A_{1g}$ symmetry}\label{subsec:a1g}

In the fully symmetric representation the leading terms in the momentum dependence of the Raman vertex are given by
Eq.~\eqref{basisfunctions} as $\gamma_0+\gamma_4\lambda_4(\mathbf{k})$, where for brevity we drop the superscript $\mu=A_{1g}$. As
pointed out before, the first term is associated with $\sigma_3$ in Nambu space, which means that its contribution to the full
$A_{1g}$ response is essentially the correlator of the charge density. However, due to the long range Coulomb interaction between
electrons light scattering on intercell charge fluctuations in the long wavelength limit is totally screened out, thus its
contribution vanishes exactly \cite{devereaux-clean}.

\begin{figure}[t]
\includegraphics[width=\columnwidth]{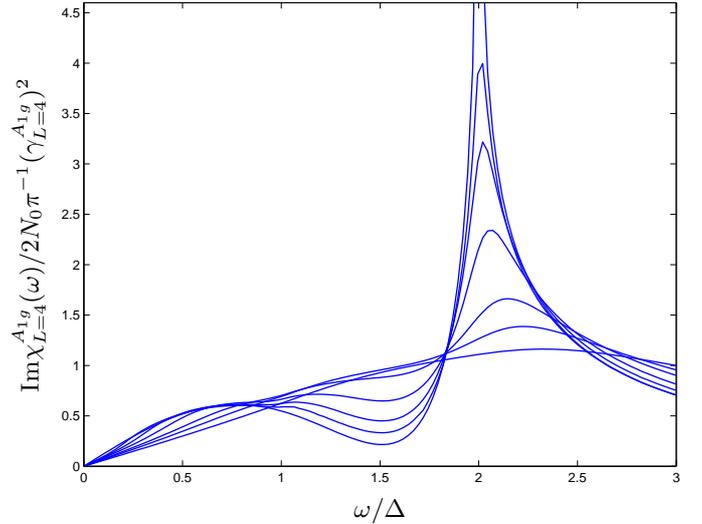}
\caption{\label{fig:a1gu}(Color online) Zero temperature $A_{1g}$ Raman spectra in dDW+dSC in the unitary limit for $\alpha=0$,
  0.025, 0.05, 0.1, 0.2, 0.3, and 0.5, from top to bottom (at $\omega=2\Delta$).}
\end{figure}

The $L=4$ term does not couple to long wavelength density fluctuations, instead it is responsible for intracell charge
fluctuations. Therefore, the total contribution to $A_{1g}$ Raman intensity comes from this term. First we have to address the
issue of vertex corrections. For this we have to solve Eq.~\eqref{vertexeq} with $M=\rho_3\sigma_3$ and
$\lambda_{L=4}(\mathbf{k})=\cos(ak_x)+\cos(ak_y)$. The solution turns out to be rather simple: vertex corrections vanish exactly
because of the mismatch between the angular dependencies of the basis function and the order parameters. From group theoretical
point of view this follows immediately from the observation that the gaps belong to $B_{1g}$, see Eqs.~\eqref{gapf} and
\eqref{basisfunctions}, whereas the basis function transforms according to a different representation \cite{devereaux-clean}. With
all these we find that the quasiparticle contribution gives the total $F$ function. In the continuum limit, where
\begin{equation}
\lambda^{A_{1g}}_{L=4}(\mathbf{k})=\cos(4\phi),\label{a1gcont}
\end{equation}
the following closed analytical formula is obtained
\begin{multline}
F(u,u')=\frac{1}{15}\frac{1}{{u'}^2-u^2}\Bigg\{\frac{E}{\sqrt{1-u^2}}
(14-18u^2+20uu'\\+28u^4-60u^3u'-24u^6+40u^5u')
+\frac{K}{\sqrt{1-u^2}}
(7u^2-15uu'\\-16u^4+40u^3u'+24u^6-40u^5u')
-[u\leftrightarrow u']\Bigg\},\label{a1gcos4phi}
\end{multline}
where the arguments of the elliptic integrals are the same as in Eq.~\eqref{f}. Substituting this expressions in the normal and
pair-breaking integrals in Eq.~\eqref{ramansusc} yields the frequency dependence of the susceptibility in the dDW+dSC state. At
zero temperature it is shown in Fig.~\ref{fig:a1gu} for different values of impurity concentration. The spectra do not depend on
the relative sizes of the dDW and dSC order parameters. The only energy scale, apart from the scattering rate of course, is the
amplitude of the effective $d$-wave gap $\Delta$. This could be inferred already from Eq.~\eqref{ramancoherence}, reflecting the
fact that Raman scattering in $A_{1g}$ symmetry is a type I process in both particle-hole and particle-particle condensates, see
Table~\ref{table:coh}. If we take a closer look on the plot we can see that there is a very small bump around $\omega=\Delta$,
which is again caused by transitions from and to the impurity band at the Fermi energy, just like in the conductivity. What is
easy to see though is that with increasing impurity concentration the peak gets suppressed and smoothed out as the lineshape
approaches the normal state result \cite{cardona}
\begin{equation}
\text{Im}\chi^{A_{1g}}_{L=4}=2N_0(\gamma^{A_{1g}}_{L=4})^2\frac{\omega\Gamma}{\omega^2+(2\Gamma)^2}.
\label{normala1g}
\end{equation}
Due to the lack of vertex corrections, here the transport lifetime equals the total lifetime, which is related to the total
scattering rate as usual via $\tau^{-1}=2\Gamma$ \cite{cardona}.

Wrapping up our discussion on the fully symmetric representation, for $\alpha=0$ the pure result is found to be
\begin{multline}\label{cleana1g}
\text{Im}\chi^{A_{1g}}_{L=4}=2N_0(\gamma^{A_{1g}}_{L=4})^2\tanh\left(\frac{\omega}{4T}\right)
\frac{1}{15}\\
\times
\begin{cases}
x^{-1}\big[(7-8x^2+16x^4)K(x) &\\
\phantom{x^{-1}\big[}-(7-12x^2+32x^4)E(x)\big] & x<1,\\
(11-28x^2+32x^4)K(x^{-1}) & \\
-(7-12x^2+32x^4)E(x^{-1}) & x>1,
\end{cases}
\end{multline}
with $x=\omega/(2\Delta)$. In the single component ground states, where either $\Delta_1=0$ or $\Delta_2=0$, it transforms
correctly to the results obtained in dSC \cite{devereaux-clean} and dDW \cite{valenzuela}.

\subsection{Raman spectra in $B_{1g}$ symmetry}\label{subsec:b1g}

Next we study Raman scattering in $B_{1g}$ symmetry. Here, the basis function coincides with the angular dependencies of the dDW
and dSC order parameters, and as such with that of the full gap as well, and for this reason we suspect that this will lead to
interference effects between the Raman vertex and the gaps \cite{devereaux-clean,raman-udw}. Restricting our analysis to the the
continuum limit as before, where
\begin{equation}
\lambda^{B_{1g}}_{L=2}(\mathbf{k})=\cos(2\phi),\label{b1gcont}
\end{equation}
the quasiparticle result is again independent of the relative sizes of $\Delta_{1,2}$, it reads
\begin{multline}
F_\text{qp}(u,u')=\frac{1}{{u'}^2-u^2}\Bigg\{\sqrt{1-u^2}\left(-uu'+\frac43+\frac{u^2}{3}\right)E\\
+\frac{u^2}{\sqrt{1-u^2}}\left(-uu'+\frac23+\frac{u^2}{3}\right)K
-[u\leftrightarrow u']\Bigg\},\label{qpb1g}
\end{multline}
but the vertex corrections do change this situation. The solution of Eq.~\eqref{vertexeq} in this symmetry channel is very lengthy
and we shall only quote here the final result. The technical details are given in Appendix~\ref{app}. The analytically
continued solution has the form
\begin{equation}
\Lambda^{B_{1g}}_{L=2}(\mathbf{k},u,u')=\cos(2\phi)\rho_3\sigma_3
+A(i\rho_1\sigma_3)+B(i\sigma_2),\label{fullb1g}
\end{equation}
where the last two terms are the corrections due to disorder. The explicit expressions for the scalar functions $A(u,u')$ and
$B(u,u')$ are given in Eqs.~\eqref{factors} and \eqref{unitaryterms}. With the impurity renormalized vertex at hand we can now go
back and evaluate Eq.~\eqref{def}. After analytically continuing to real frequencies we find the general expression in
Eq.~\eqref{ramansusc} with the total $F$ function
\begin{equation}
F(u,u')=F_\text{qp}(u,u')+\frac{\Delta_1}{\Delta}AQ+\frac{\Delta_2}{\Delta}BQ,\label{fullfb1g}
\end{equation}
where $Q(u,u')$ is given explicitly in Eq.~\eqref{q2} in Appendix.

\begin{figure}[t]
\includegraphics[width=\columnwidth]{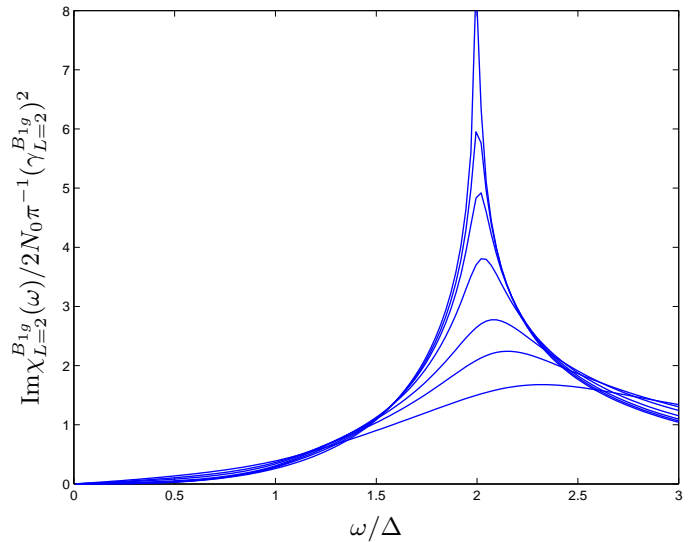}
\caption{\label{fig:b1gu}(Color online) Zero temperature $B_{1g}$ Raman spectra in dDW+dSC in the unitary limit for $\alpha=0$,
  0.025, 0.05, 0.1, 0.2, 0.3, and 0.5 from top to bottom (at $\omega=2\Delta$), and $\Delta_1/\Delta=2/3$.}
\end{figure}

This is the point where we can explicitly see what we have conjectured in the beginning of this subsection. Namely, the
interference of the angular dependencies of the bare vertex and the order parameters resulted in additional terms in the Raman
susceptibility. If the ground state is dDW alone, where $\Delta_2=0$, the contribution from ladder diagrams is $AQ$. In dSC the
roles simply interchange and we find $BQ$. In general when both orders contribute to the coexisting phase neither of these
correction vanish. The $B_{1g}$ Raman spectra in the dDW+dSC ground state is shown in Fig.~\ref{fig:b1gu}. We can observe that the
peak at $2\Delta$ dominates the spectra. This is in fact not that surprising, because the type I coherence factor is so, as
opposed to the conductivity, that it facilitates this transition in both the dDW and dSC components of the coexisting phase, thus
making it more pronounced. With increasing impurity concentration the cubic low frequency behavior of the pure system changes to
linear tendency, the sharp peak widens and looses of its intensity as the lineshapes approach the normal state result
\cite{cardona}
\begin{equation}
\text{Im}\chi^{B_{1g}}_{L=2}=2N_0(\gamma^{B_{1g}}_{L=2})^2\frac{\omega\Gamma}{\omega^2+(2\Gamma)^2}.
\label{normalb1g}
\end{equation}
Apart from trivial substitutions of indices this is precisely the same as Eq.~\eqref{normala1g}, showing that in an isotropic
metal the different symmetry spectra cannot be distinguished. There is a remark though we would like to make at this point. We see
that it is again, interestingly, the total lifetime of electrons $\tau^{-1}=2\Gamma$ showing up in Eq.~\eqref{normalb1g}, and not
the transport lifetime, although there are vertex corrections in this symmetry channel. The reason lies in the fact that in our
approximation the scattering potential $U$, see Eq.~\eqref{potential}, does not depend on the momenta of incoming and scattered
particles. Hence its expansion in the basis functions of the point group consists of only the momentum independent $L=0$ term,
which cannot change the lifetime in the $L=2$ channel \cite{cardona}.

Again, we finish this subsection by stating the pure result in dDW+dSC, for which closed analytical expression exists
\begin{multline}\label{cleanb1g}
\text{Im}\chi^{B_{1g}}_{L=2}=2N_0(\gamma^{B_{1g}}_{L=2})^2\tanh\left(\frac{\omega}{4T}\right)
\frac{1}{3}\\
\times
\begin{cases}
x^{-1}\left((2+x^2)K(x)-2(1+x^2)E(x)\right) & x<1,\\
(1+2x^2)K(x^{-1})-2(1+x^2)E(x^{-1}) & x>1,
\end{cases}
\end{multline}
where $x=\omega/(2\Delta)$. If we are on the phase diagram where the system does not support dDW ordering and only $d$-wave
superconductivity exists, we get back correctly the known result of dSC \cite{devereaux-clean}. The reverse statement is also true:
for $\Delta=\Delta_1$ the above expression yields the dDW result as well \cite{valenzuela}.

\subsection{Raman spectra in $B_{2g}$ symmetry}\label{subsec:b2g}

Last but not least let us summarize the results obtained for $B_{2g}$ symmetry. Here the basis function is $\sin(ak_x)\sin(ak_y)$,
and the Nambu matrix appearing in the effective density is $M=\sigma_3$, resembling the true charge density. Consequently, in dDW
Raman scattering in this symmetry is a type II process, see Table~\ref{table:coh}. Quite similarly to what we have discussed in
$A_{1g}$ symmetry, one finds that the vertex is not renormalized by impurities because of the mismatch between the basis function
and the $d$-wave order parameters. They belong to different representations, hence the orthogonality. Also, it is worth mentioning
that though the vertex possesses the same Nambu structure as the density itself, still there is no Coulomb screening involved,
because the $L=0$ and $L=2$ basis functions are orthogonal too. This leads to the vanishing of the cross-susceptibility
$\langle[\sigma_3,\gamma_2\lambda_2\sigma_3]\rangle$, so the screening RPA series cannot build up. From physical point of view the
reason is that in case of short wavelength intracell density fluctuations the Coulomb forces are irrelevant
\cite{devereaux-clean}.

\begin{figure}[t]
\includegraphics[width=\columnwidth]{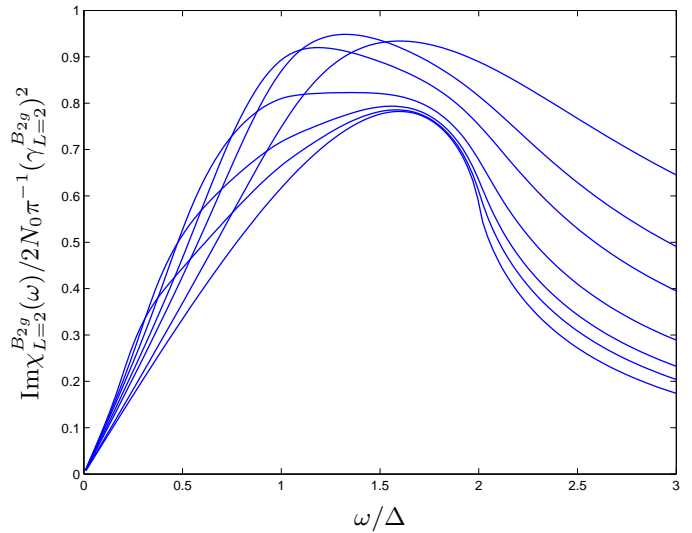}
\caption{\label{fig:b2gu}(Color online) Zero temperature $B_{2g}$ Raman spectra in dDW+dSC in the unitary limit for $\alpha=0$,
  0.025, 0.05, 0.1, 0.2, 0.3, and 0.5 from bottom to top (at $\omega=3\Delta$), and $\Delta_1/\Delta=2/3$.}
\end{figure}

In the continuum limit, where the basis function simplifies to
\begin{equation}
\lambda^{B_{2g}}_{L=2}(\mathbf{k})=\sin(2\phi),\label{b2gcont}
\end{equation}
the total $F$ function is again obtained from the one bubble calculation, from Eq.~\eqref{ramancoherence}. Due to the opposite
coherence factors of dDW and dSC even the quasiparticle result shows explicit dependence on the relative sizes of the gaps
\begin{multline}
F(u,u')=\frac{1}{{u'}^2-u^2}\Bigg\{\sqrt{1-u^2}\Bigg(u(u+u')(E-K)\\
+\frac23\left(\frac{\Delta_2}{\Delta}\right)^2
((1-2u^2)E+2u^2K)\Bigg)-[u\leftrightarrow u']\Bigg\},\label{qpb2g}
\end{multline}

where, again, the argument of $K$ and $E$ is the same as in Eq.~\eqref{f}. The spectra in the dDW+dSC ground state are shown in
Fig.~\ref{fig:b2gu}. The main qualitative difference here compared to the other spectra is that the obvious peak at $2\Delta$ is
missing. This is again due to the mismatch between the angular dependencies of the bare vertex and the gap functions, the same
effect that lead to the vanishing of vertex corrections. With increasing pair-breaking the curves smoothly approach the metallic
result \cite{cardona}, just as they do in the other representations
\begin{equation}
\text{Im}\chi^{B_{2g}}_{L=2}=2N_0(\gamma^{B_{2g}}_{L=2})^2\frac{\omega\Gamma}{\omega^2+(2\Gamma)^2}.
\label{normalb2g}
\end{equation}

The pure result for $\alpha=0$ can again be expressed in closed form
\begin{multline}\label{cleanb2g}
\text{Im}\chi^{B_{2g}}_{L=2}=2N_0(\gamma^{B_{2g}}_{L=2})^2\left(\frac{\Delta_2}{\Delta}\right)^2
\tanh\left(\frac{\omega}{4T}\right)\frac{1}{3}\\
\times
\begin{cases}
x^{-1}\left((1-x^2)K(x)-(1-2x^2)E(x)\right) & x<1,\\
2(1-x^2)K(x^{-1})-(1-2x^2)E(x^{-1}) & x>1,
\end{cases}
\end{multline}
with $x=\omega/(2\Delta)$. As expected, in a single component dSC with $\Delta=\Delta_2$ we reobtain the known result and
lineshape \cite{devereaux-clean}. However, the other extreme where only dDW exists is rather interesting and somewhat controversial
as far as the literature is concerned. In pure system Eq.~\eqref{cleanb2g} reveals that because of the type II behavior there is
no absorption at any frequency in dDW. This result is completely analogue to the optical conductivity of pure superconductors,
which too vanishes in the long wavelength limit, again because of the type II coherence factor \cite{mahan-book}. To further deepen
the analogy we find that in the presence of non-magnetic impurities, exactly as in the case of the infrared response in
superconductors \cite{skalski,inplane-dsc,mattis}, the Raman intensity becomes finite, but is certainly very small. This result of
ours is in precise agreement with the findings of Ref.~\cite{zeyher-greco}, which, though focuses on clean cuprate
superconductivity, comes to the conclusion that in the strongly underdoped region of the phase diagram the $B_{2g}$ Raman response
is very small due to the relation $\Delta_2\ll\Delta_1$. We must note at this point, however, that other authors \cite{valenzuela}
find finite $B_{2g}$ Raman susceptibility in pure dDW, which in our opinion is probably not correct. The discrepancy in their
calculation originates most probably from the fact that they derived the Raman susceptibility from the conductivity, and might
have overlooked the different Nambu structure of the current operator $(\rho_3)$ and the effective density $(\rho_0=1)$ in this
symmetry.

\section{Conclusions}\label{sec:conclusions}

In summary we have studied coherence effects and disorder in the ground state of a quasi two-dimensional coexisting $d$-wave
density wave and $d$-wave superconductor (dDW+dSC). The main purpose of the paper was to generalize and extend available
calculations on in-plane optical conductivity and electronic Raman response in the sense to incorporate impurity scattering on a
microscopic level and to allow for the possibility of a coexisting phase with two order parameters, especially relevant in
underdoped cuprates.

We calculated the frequency dependence of the in-plane optical conductivity and the superfluid density in dDW+dSC in the presence
of impurities in unitary limit. The dDW component of the ground state results in finite absorption at all frequencies, even in the
pure system, and in particular in a strong peak around twice the amplitude of the effective $d$-wave gap, which is reminiscent of
the type I coherence behavior known from density wave theory. As to the superfluid density, we found that due to the transfer of
spectral weight from zero to finite frequency, not even in the pure system would it reach unity, instead it is reduced according
to the the very simple expression: $1-\Delta_1^2/\Delta^2$.

We determined the electronic Raman susceptibility in dDW+dSC. The calculations were carried out for the $A_{1g}$, $B_{1g}$ and
$B_{2g}$ symmetries. In accordance with available results on Raman lineshapes of cuprates, the spectra are strongly channel
dependent. The qualitative differences, however, continuously disappear with increasing impurity concentration as the curves
approach the Lorentz-like metallic form. We found that the dDW and dSC components of the coexisting dDW+dSC state contribute
practically the same in $A_{1g}$ and $B_{1g}$ symmetries, because here the coherence factors in both condensates are type I. As
opposed to this, in $B_{2g}$ the Nambu structure of the Raman vertex changes from $\rho_3\sigma_3$ to $\sigma_3$, which at the
same time changes the dDW coherence factor too from type I to type II. This in turn leads to the interesting conclusion that the
Raman intensity in this channel is dominated by the dSC contribution, the dDW part survives only in dirty materials.

\begin{acknowledgement}
This work was supported by the Hungarian National Research Fund under Grant No.\ OTKA K 72613. One of us (A.\ V\'anyolos) would
like to thank the hospitality of the Max-Planck-Institut f\"ur Komplexer Systeme in Dresden (MPIPKS), where a part of this work
was done.
\end{acknowledgement}

\appendix
\section{Vertex corrections to $B_{1g}$ Raman vertex}\label{app}

In this appendix we summarize the technical details of the solution of the vertex equation for $B_{1g}$
symmetry. The calculation is carried out in the unitary limit. In this symmetry $M=\rho_3\sigma_3$, and the bare
vertex in the continuum limit takes the form
\begin{equation}
\lambda^{B_{1g}}_{L=2}(\mathbf{k})=\cos(2\phi).\label{bare}
\end{equation}
In unitary limit the $T$-matrix is given by Eq.~\eqref{tunitary}. Direct substitution of the bare vertex into the right hand side
of Eq.~\eqref{vertexeq} shows that the $\rho_3\sigma_3$ matrix structure cannot be a consistent solution alone, because due to the
interference of $\cos(2\phi)$ with the angular dependencies of $\Delta_{1,2}(\mathbf{k})$, new terms appear, proportional to
$\rho_1\sigma_3$ and $\sigma_2$. Therefore, the first iteration already suggests that we have to allow for the existence of new
terms in Nambu space, and this leads us to the ansatz
\begin{equation}
\Lambda^{B_{1g}}_{L=2}(\mathbf{k},u,u')=\cos(2\phi)\rho_3\sigma_3+A(i\rho_1\sigma_3)+B(i\sigma_2).
\label{fullvertexb1g}
\end{equation}
Substituting this into both sides of Eq.~\eqref{vertexeq} one finds that this is indeed a consistent solution, and the
problem is now reduced to finding the solution of two coupled linear equations for the unknown coefficients $A$ and $B$. For these
in turn we get
\begin{subequations}\label{factors}
\begin{align}
A&=\frac{be-a(f-1)}{(c-1)(f-1)-de},\label{A}\\
B&=\frac{ad-b(c-1)}{(c-1)(f-1)-de},\label{B}
\end{align}
\end{subequations}
where
\begin{subequations}\label{unitaryterms}
\begin{align}
a&=\frac{\pi\Gamma}{2\Delta}\frac{\sqrt{1-u^2}\sqrt{1-{u'}^2}}{uu'KK'}\frac{\Delta_1}{\Delta}Q(u,u'),\label{aunitary}\\
b&=\frac{\pi\Gamma}{2\Delta}\frac{\sqrt{1-u^2}\sqrt{1-{u'}^2}}{uu'KK'}\frac{\Delta_2}{\Delta}Q(u,u'),\label{bunitary}
\end{align}
\begin{multline}
c=\frac{\pi\Gamma}{2\Delta}\frac{\sqrt{1-u^2}\sqrt{1-{u'}^2}}{uu'KK'}\frac{1}{u'-u}
\Bigg(\frac{u-u'}{u+u'}\\
\times\left(\frac{uK}{\sqrt{1-u^2}}+\frac{u'K'}{\sqrt{1-{u'}^2}}\right)
+2Q\left(\frac{\Delta_2}{\Delta}\right)^2\Bigg),\label{cunitary}
\end{multline}
\begin{align}
d&=\frac{\pi\Gamma}{\Delta}\frac{\sqrt{1-u^2}\sqrt{1-{u'}^2}}{uu'KK'}\frac{1}{u-u'}\frac{\Delta_1\Delta_2}{\Delta^2}
Q(u,u'),\label{dunitary}\\
e&=d,\label{eunitary}
\end{align}
\begin{multline}
f=\frac{\pi\Gamma}{2\Delta}\frac{\sqrt{1-u^2}\sqrt{1-{u'}^2}}{uu'KK'}\frac{1}{u'-u}
\Bigg(\frac{u-u'}{u+u'}\\
\times\left(\frac{uK}{\sqrt{1-u^2}}+\frac{u'K'}{\sqrt{1-{u'}^2}}\right)
+2Q\left(\frac{\Delta_1}{\Delta}\right)^2\Bigg),\label{funitary}
\end{multline}
and
\begin{multline}
Q=\frac{1}{u+u'}\Bigg(\sqrt{1-{u'}^2}E'+\frac{{u'}^2K'}{\sqrt{1-{u'}^2}}\\
-\sqrt{1-u^2}E-\frac{u^2K}{\sqrt{1-u^2}}\Bigg).\label{q2}
\end{multline}
\end{subequations}
In these expressions the argument of $K$ and $E$, the elliptic integrals of the first and second kind, is
$1/\sqrt{1-u^2}$, whereas in $K'$ and $E'$ it is $1/\sqrt{1-u'^2}$.

\bibliographystyle{epj}
\bibliography{paper}

\begin{thebibliography}{58}

\bibitem{laughlin}
S.~Chakravarty, R.B. Laughlin, D.K. Morr, C.~Nayak, Phys. Rev. B \textbf{63},
  094503 (2001)

\bibitem{emery}
V.J. Emery, S.A. Kivelson, Nature \textbf{374}, 434 (1995)

\bibitem{benfatto}
L.~Benfatto, S.~Caprara, C.~{Di Castro}, Eur. Phys. J. B \textbf{17}, 95 (2000)

\bibitem{zeyher-greco}
R.~Zeyher, A.~Greco, Phys. Rev. Lett. \textbf{89}, 177004 (2002)

\bibitem{valenzuela}
B.~Valenzuela, E.J. Nicol, J.P. Carbotte, Phys. Rev. B \textbf{71}, 134503
  (2005)

\bibitem{ding}
H.~Ding, T.~Yokoya, J.C. Campuzano, T.~Takahashi, M.~Randeria, M.R. Norman,
  T.~Mochiku, K.~Kadowaki, J.~Giapintzakis, Nature \textbf{382}, 51 (1996)

\bibitem{loeser}
A.G. Loeser, Z.X. Shen, D.S. Dessau, D.S. Marshall, C.H. Park, P.~Fournier,
  A.~Kapitulnik, Science \textbf{273}, 325 (1996)

\bibitem{norman}
M.R. Norman, H.~Ding, M.~Randeria, J.C. Campuzano, T.~Yokoya, T.~Takeuchi,
  T.~Takahashi, T.~Mochiku, K.~Kadowaki, P.~Guptasarma et~al., Nature
  \textbf{392}, 157 (1998)

\bibitem{renner}
C.~Renner, B.~Revaz, J.Y. Genoud, K.~Kadowaki, {\O}.~Fischer, Phys. Rev. Lett.
  \textbf{80}, 149 (1998)

\bibitem{kim}
W.~Kim, J.X. Zhu, J.P. Carbotte, C.S. Ting, Phys. Rev. B \textbf{65}, 064502
  (2002)

\bibitem{congjun1}
C.~Wu, W.V. Liu, Phys. Rev. B \textbf{66}, 020511 (2002)

\bibitem{bena}
C.~Bena, S.~Chakravarty, J.~Hu, C.~Nayak, Phys. Rev. B \textbf{69}, 134517
  (2004)

\bibitem{morr}
D.K. Morr, Phys. Rev. Lett. \textbf{89}, 106401 (2002)

\bibitem{jian}
J.X. Zhu, W.~Kim, C.S. Ting, J.P. Carbotte, Phys. Rev. Lett. \textbf{87},
  197001 (2001)

\bibitem{marston}
J.P. Marston, J.O. Fj\ae{restad}, Phys. Rev. Lett. \textbf{89}, 056404 (2002)

\bibitem{ismer-spin}
J.P. Ismer, I.~Eremin, D.K. Morr, Phys. Rev. B \textbf{73}, 104519 (2006)

\bibitem{ghosh-coexistance}
H.~Ghosh, A.~Singh, Phys. Rev. B \textbf{66}, 064530 (2002)

\bibitem{andrenacci}
N.~Andrenacci, G.G.N. Angilella, H.~Beck, R.~Pucci, Phys. Rev. B \textbf{70},
  024507 (2004)

\bibitem{gallais}
Y.~Gallais, A.~Sacuto, T.P. Devereaux, D.~Colson, Phys. Rev. B \textbf{71},
  012506 (2005)

\bibitem{yu}
L.~Yu, D.~Munzar, A.V. Boris, P.~Yordanov, J.~Chaloupka, T.~Wolf, C.T. Lin,
  B.~Keimer, C.~Bernhard, Phys. Rev. Lett. \textbf{100}, 177004 (2008)

\bibitem{pan}
S.H. Pan, E.W. Hudson, K.M. Lang, H.~Eisaki, S.~Uchida, J.C. Davis, Nature
  \textbf{403}, 746 (2000)

\bibitem{hotta}
T.~Hotta, J. Phys. Soc. Jpn. \textbf{62}, 274 (1993)

\bibitem{pssb-gossamer}
H.~Won, S.~Haas, K.~Maki, D.~Parker, B.~D\'ora, A.~Virosztek, Phys. Stat. Sol.
  (b) \textbf{243}, 37 (2006)

\bibitem{pssb-bottomup}
H.~Won, S.~Haas, K.~Maki, Phys. Stat. Sol. (b) \textbf{244}, 2407 (2007)

\bibitem{pssb-hightc}
H.~Won, Y.~Morita, K.~Maki, Phys. Stat. Sol. (b) \textbf{244}, 4371 (2007)

\bibitem{abrikosov}
A.A. Abrikosov, L.P. Gor'kov, Zh. Eksp. Teor. Fiz \textbf{39}, 1781 (1960),
  [Sov. Phys. JETP \textbf{12}, 1243 (1961)]

\bibitem{skalski}
S.~Skalski, O.~Betbeder-Matibet, P.R. Weiss, Phys. Rev. \textbf{136}, A1500
  (1964)

\bibitem{ambegaokar}
V.~Ambegaokar, A.~Griffin, Phys. Rev. \textbf{137}, A1151 (1965)

\bibitem{balazs-born}
B.~D\'ora, A.~Virosztek, K.~Maki, Phys. Rev. B \textbf{66}, 115112 (2002)

\bibitem{balazs-unitary}
B.~D\'ora, A.~Virosztek, K.~Maki, Phys. Rev. B \textbf{68}, 075104 (2003)

\bibitem{nca-impurity}
A.~V\'anyolos, B.~D\'ora, K.~Maki, A.~Virosztek, New J. Phys. \textbf{9}, 216
  (2007)

\bibitem{maki-dwave}
K.~Maki, \emph{Introduction to {$d$}-wave superconductivity}, in \emph{Lectures
  on the {P}hysics of {H}ighly {C}orrelated {E}lectron {S}ystems}, edited by
  F.~Mancini (AIP, Woodbury, New York, 1998), Vol. 438, pp. 83--128

\bibitem{maki-nodal}
H.~Won, S.~Haas, D.~Parker, S.~Telang, A.~V\'anyolos, K.~Maki, \emph{{BCS}
  theory of nodal superconductors}, in \emph{Lectures on the {P}hysics of
  {H}ighly {C}orrelated {E}lectron {S}ystems {IX}}, edited by A.~Avella,
  F.~Mancini (AIP, Melville, New York, 2005), Vol. 789, pp. 3--43

\bibitem{abrikosov-book}
A.A. Abrikosov, L.P. Gor'kov, I.E. Dzyaloshinski, \emph{Methods of Quantum
  Field Theory in Statistical Physics} (Dover, New York, 1963)

\bibitem{rickayzen-book}
G.~Rickayzen, \emph{Green's Functions and Condensed Matter} (Academic Press,
  London, 1980)

\bibitem{parks-book}
K.~Maki, in \emph{Superconductivity}, edited by R.D. Parks (Marcel Dekker, New
  York, 1969), Vol.~2, pp. 1035--1105

\bibitem{sun-maki}
Y.~Sun, K.~Maki, Phys. Rev. B \textbf{51}, 6059 (1995)

\bibitem{puch-maki}
E.~Puchkaryov, K.~Maki, Eur. Phys. J. B \textbf{4}, 191 (1998)

\bibitem{jiang-inelastic}
C.~Jiang, J.P. Carbotte, Phys. Rev. B \textbf{53}, 11868 (1996)

\bibitem{devereaux-clean}
T.P. Devereaux, D.~Einzel, Phys. Rev. B \textbf{51}, 16336 (1995)

\bibitem{inplane-dsc}
B.~D\'ora, K.~Maki, A.~Virosztek, Curr. Appl. Phys. \textbf{6}, 903 (2006)

\bibitem{outofplane-dsc}
B.~D\'ora, K.~Maki, A.~Virosztek, Europhys. Lett. \textbf{55}, 847 (2001)

\bibitem{mahan-book}
G.D. Mahan, \emph{Many-{P}article {P}hysics} (Kluwer Academic/Plenum
  Publishers, 233 Spring Street, New York 10013, 2000)

\bibitem{tinkham-book}
M.~Tinkham, \emph{Introduction to Superconductivity} (McGraw-Hill, New York,
  1996)

\bibitem{fetter-book}
A.L. Fetter, J.D. Walecka, \emph{Quantum {T}heory of {M}any-{P}article
  {S}ystems} (Dover Publications, 31 East Second Street Mineola, New York
  11501, 2003)

\bibitem{impurity-sdw}
A.~Virosztek, B.~D\'ora, K.~Maki, Europhys. Lett. \textbf{47}, 358 (1999)

\bibitem{lee}
P.A. Lee, Phys. Rev. Lett. \textbf{71}, 1887 (1993)

\bibitem{balazs-sdw}
B.~D\'ora, A.~Virosztek, Eur. Phys. J. B \textbf{22}, 167 (2001)

\bibitem{lra}
P.A. Lee, T.M. Rice, P.W. Anderson, Solid State Commun. \textbf{14}, 703 (1974)

\bibitem{phonon-ucdw}
A.~V\'anyolos, B.~D\'ora, A.~Virosztek, Phys. Rev. B \textbf{73}, 165127 (2006)

\bibitem{dierker}
M.V. Klein, S.B. Dierker, Phys. Rev. B \textbf{29}, 4976 (1984)

\bibitem{devereaux-rmp}
T.~Devereaux, R.~Hackl, Rev. Mod. Phys. \textbf{79}, 175 (2007)

\bibitem{tutto-cdw-sc}
I.~T\"utt\H{o}, A.~Zawadowski, Phys. Rev. B \textbf{45}, 4842 (1992)

\bibitem{raman-udw}
A.~V\'anyolos, A.~Virosztek, Phys. Rev. B \textbf{72}, 115119 (2005)

\bibitem{sigrist}
M.~Sigrist, K.~Ueda, Rev. Mod. Phys. \textbf{63}, 239 (1991)

\bibitem{scalapino}
D.J. Scalapino, Phys. Rep. \textbf{250}, 329 (1995)

\bibitem{cardona}
A.~Zawadowski, M.~Cardona, Phys. Rev. B \textbf{42}, 10732 (1990)

\bibitem{mattis}
D.C. Mattis, J.~Bardeen, Phys. Rev. \textbf{111}, 412 (1958)

\end{thebibliography}

\end{document}